\documentclass[twocolumn]{revtex4-1}
\usepackage{graphicx}
\usepackage{epstopdf}
\usepackage{amsmath}
\begin{document}
\title{Formation of deuterons by coalescence:\\ Consequences on the deuteron
number fluctuations}
\author{Zuzana Feckov\' a$^{a,b}$}
\author{Jan Steinheimer$^{c}$}
\author{Boris Tom\'a\v{s}ik$^{b,d}$}
\author{Marcus Bleicher$^{c,e}$}
\affiliation{$^a$Univerzita Pavla Jozefa \v{S}af\'arika, 
\v{S}rob\'{a}rova 2, 04001 Ko\v{s}ice, Slovakia}
\affiliation{$^b$Univerzita Mateja Bela, Tajovsk\'eho 40, 97401 Bansk\'a Bystrica, Slovakia}
\affiliation{$^c$Frankfurt Institude for Advanced Studies, 
Johann Wolfgang Goethe-Universit\"at, Ruth-Moufang-Strasse 1, 60438 Frankfurt am Main}
\affiliation{$^d$\v{C}esk\'e vysok\'e u\v{c}en\'i technick\'e v Praze, 
FJFI, B\v{r}ehov\'a 7, 11519 Praha 1, Czech Republic}
\affiliation{$^e$Institut f\"ur Theoretische Physik, 
Johann Wolfgang Goethe-Universit\"at, Max-von-Laue-Strasse 1, 60438 Frankfurt am Main}

\begin{abstract}
Two scenarios for cluster production have since long been discussed in the literature: i) direct emission of the clusters from a (grand canonical) thermal source or ii) subsequent formation of the clusters by coalescence of single nucleons. While both approaches have been successfully applied in the past  it has not yet been clarified which of the two mechanisms dominates the cluster production. 
We propose to use recently developed event-by-event techniques 
to study particle multiplicity fluctuations
on nuclear clusters and employ this analysis to the deuteron
 number fluctuations to disentangle the two production mechanisms. We argue that for a grand canonical cluster formation, the cluster fluctuations will follow  Poisson distribution, 
while for the coalescence scenario, the fluctuations will strongly deviate from the 
Poisson expectation.
We estimate the effect to be 10\% for the variance and up to a factor of 5 for the 
kurtosis of the deuteron number multiplicity distribution. Our prediction can be tested in the 
beam energy scan program at 
RHIC as well as experiments  at the FAIR and NICA facilities.
\end{abstract}
\maketitle


\section{Introduction}
The formation of bound states in nuclear collisions has been investigated experimentally and theoretically for many decades. On the experimental side the fragmentation of the source has been studied extensively at very low energies to extract information about the nuclear liquid-gas phase transition \cite{Panagiotou:1984rb,Pochodzalla:1993ex}, while on the theoretical side those observations were studied in detail by \cite{Csernai:1987xz,Pratt:1987zz,Aichelin:1988me,Aichelin:1991xy}. Already very early it was pointed out that deuteron production can be used to infer thermodynamic properties of the system, e.g. the entropy
per baryon
\cite{Kapusta:1984ij}. At higher energies around $\sqrt{s_{NN}}=3-20$ GeV, the data become more scarce. Here, the AGS experiments have provided data on the formation of clusters up to Helium \cite{Wang:1994rua}, while the SPS experiments have measured data on clusters up to mass three \cite{Bearden:1999iq,Baatar:2012fua} even for the anti-particle sector \cite{Weber:2002xr}. Currently the RHIC and LHC experiments have also shown data on light (anti-)nuclei production at the highest energies, see e.g. \cite{Adler:2001uy,Adam:2015vda}.  Theoretically, the mechanism for cluster production at intermediate and high energies is not well understood. It is clear that clusters do not stem from a break-up of the initial 
target and projectile 
nuclei but have to be formed newly towards the end of the fireball evolution. However, it is not a priori clear, if the clusters are directly formed at the chemical freeze-out, 
e.g. from a grand canonical thermal ensemble \cite{Andronic:2010qu}, 
or if the formation of the clusters happens at the kinetic 
surface by coalescence of individual nucleons \cite{Nagle:1996vp}. While one may argue that lightly bound clusters (e.g deuteron has a binding energy of only a few MeV) may either not be formed at the chemical freeze-out due to the large temperature and may be easily destroyed (if formed earlier) in the kinetic stage, the predictions of clusters multiplicities within the statistical approach provide a good description of the measured data. On the other hand coalescence of neutrons and protons to deuterons after the kinetic freeze-out is certainly another possible process and allows to describe the experimental data equally well \cite{Nagle:1996vp}.

In this paper we propose to use the fluctuations of the deuteron number to distinguish between the two production/formation mechanisms. These studies have become possible due to the increased experimental possibilities available at RHIC BES.


\section{Set up}
To elucidate the idea we compare three scenarios: 
i) Direct deuteron production from a grand canonical thermal ensemble at the chemical 
freeze-out. Here, all fluctuations are Poissonian and the scaled moments of the deuteron distribution $\sigma^2/\lambda$, $S\sigma$, and $\kappa\sigma^2$ are all unity. 
In addition, there is no correlation between the proton and the deuteron number. 
ii) Only production of nucleons from 
the grand canonical thermal ensemble at the chemical freeze-out is assumed, 
while deuterons are formed by coalescence after the kinetic freeze-out. In this case 
we consider two variations of the coalescence prescription: 
ii.a)  The initial number of protons fluctuates according to a Poisson distribution, 
while the number of deuterons depends on the squared proton density. 
ii.b) Both proton and neutron fluctuations are Poissonian and the number of 
deuterons is proportional to their product. 
As a result in both coalescence scenarios the deuteron number will not fluctuate 
according to a Poisson distribution and all higher moments will show strong 
deviations from their Poisson values. 

Next we closer explain both used coalescence models. 


\subsection{Model A: Correlated proton and neutron number}

To quantify the fluctuations in scenario ii.a) we follow the standard 
coalescence approach, i.e. deuterons are formed  after the kinetic 
freeze-out in each event with a probability proportional to the squared number 
of all initially produced protons \cite{Butler:1963pp,Nagle:1996vp}, i.e.
\begin{equation}
\label{coal}
\lambda_d = B n_i^2, 
\end{equation}
where $B$ denotes the coalescence parameter, which may depend on the center-of-mass energy. In this model we actually make the (standard) assumption that neutron 
yield is directly correlated with the proton yield in that event. (This assumption 
will be relaxed in the other model.)
The strength of the event-by-event proton-neutron correlation can be extracted from experimental measurements, see also Appendix.
We emphasize that the predicted fluctuation signal should be studied in a fixed volume, e.g. using tight centrality cuts to avoid volume fluctuations. 

The number of deuterons $n_d$ in a given reaction with fixed initial proton number $n_i$ is then given by a Poisson distribution
\begin{equation}
P_d(n_d|n_i)=\lambda_d^{n_d} \frac{e^{-\lambda_d}}{n_d!} =
\left ( B n_i^2 \right )^{n_d} \frac{e^{-B n_i^2}}{n_d!} 
\, .
\end{equation} 
Summing over the initial proton numbers distributed according to $P_i(n_i)$ 
then leads to the final deuteron number fluctuations based on the initial 
proton number fluctuations as 
\begin{equation}
P_d(n_d) = \sum_{n_i\geq n_d} P_d(n_d|n_i) P_i(n_i).
\label{e:ddist}
\end{equation}
Recall that $P_i(n_i)$ is Poissonian and the initial proton number can be obtained as
\begin{equation}
n_i = n_p + n_d
\end{equation}
where $n_p$ is the mean \emph{observed} proton number.


\subsection{Model B: Independent proton and neutron fluctuations}

The coalescence model presented in previous subsection
is extreme in its assumption that neutron and proton fluctuations are 
strongly correlated. In order to make robust predictions for the moments 
of the deuteron number distribution we will relax the assumption completely 
and consider the model with independent proton and neutron fluctuations. The 
initial proton 
(neutron) number $n_i$($n_j$) fluctuates again according to Poisson distribution 
and the deuteron formation probability is proportional to the product of nucleon 
multiplicities, i.e.
\begin{equation}
\label{e:lamb}
\lambda_d = B n_i n_j.
\end{equation}
The coalescence parameter $B$ depends again only on the collision energy.
The initial neutron number $n_j$ fluctuates according to a Poisson distribution with the 
same mean number as the initial proton number. 

The number of deuterons in events with given number of nucleons is given by the 
Poisson distribution
\begin{equation}
P_d(n_d|n_i,n_j)=\lambda_d^{n_d} \frac{e^{-\lambda_d}}{n_d!} =
\left ( B n_i n_j \right )^{n_d} \frac{e^{-B n_i n_j}}{n_d!} 
\, .
\end{equation} 
The deuteron number distribution is then obtained by summing up over the initial proton and neutron number distributions
\begin{equation}
P_d(n_d) = \sum_{n_i,n_j\geq n_d} P_d(n_d|n_i,n_j) P_i(n_i) P_j(n_j).
\end{equation}

It has been shown recently  that if deuteron production scales with the 
squared proton number (Model A) then higher scaled moments 
of the observed proton number 
fluctuations agree qualitatively with the data from RHIC beam energy scan
program \cite{Aggarwal:2010wy,Feckova:2015qza}. Unfortunetely, this feature does 
not survive in Model B. Nevertheless, we also use it in our calculation. The real 
situation is perhaps somewhere in between the two used models and by showing that 
our signal is clearly visible in both cases we will prove its robustness. 


\section{Moments of the deuteron distribution}

\begin{figure}[t]
\includegraphics[width=0.5\textwidth]{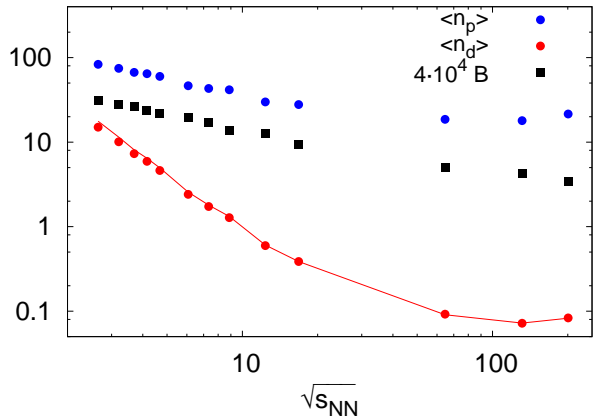}
\caption{\label{fig:input}  Model parameter $B$ for Model A (black squares) and resulting proton (in blue) and deuteron (in red) multiplicities as function of energy. The resulting deuteron multiplicity is compared to the thermal fit (red line) input to our model.}
\end{figure}

The essential model parameter is 
the coalescence parameter $B$ which we fix to obtain the correct mean deuteron multiplicity at midrapidity for each energy. While the data on proton multiplicities at midrapidity are abundant, the data on deuterons are not available for all the examined energies. The available data are however well reproduced by the thermal model. Deuteron to proton ratio $d/p$ can be thus parametrized as:
\begin{equation}
\frac{d}{p} = 0.8 \left [ \frac{\sqrt{s_{NN}}}{1\, \mathrm{GeV}} \right ]^{-1.55} + 0.0036\, .
\end{equation}
The observed proton and parametrized deuteron multiplicites are well reproduced in our model. We summarize the model parameter $B$ and results for the mean values of proton and deuteron multiplicities for various collision energies in Fig.~\ref{fig:input}. 
Plotted are values for Model A. The difference to Model B is so small that if it was also 
plotted the data points would practically overlap.

\begin{figure}[t]
\includegraphics[width=0.5\textwidth]{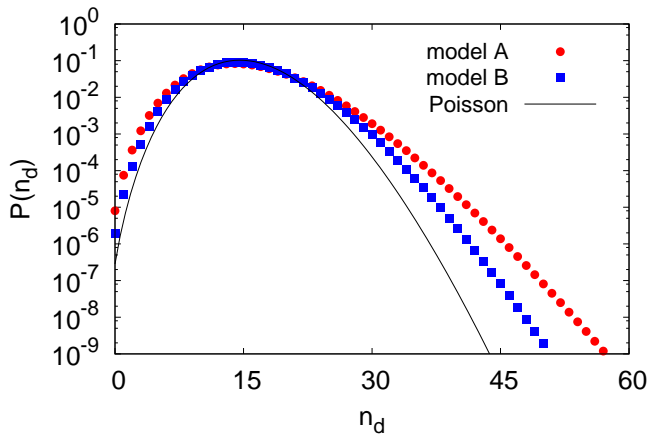}
\caption{\label{fig:deut}  Fluctuation of the deuteron number for Au+Au collisions at 2.6 GeV beam energy in comparison to the Poisson distribution. The parameters of the distributions 
are for Model A: $\sigma^2/\langle n_d \rangle = 1.609$, $S\sigma = 2.218$, $\kappa \sigma^2 = 6.915$; 
Model B:  $\sigma^2/\langle n_d \rangle = 1.308$, $S\sigma = 1.616$, $\kappa \sigma^2 = 3.422$.}
\end{figure} 
For illustration,  
Fig.~\ref{fig:deut} shows the distributions of  deuteron number for Au+Au collisions at 
2.6 GeV beam energy in comparison to the Poisson distribution. Here one clearly observes 
that coalescence leads to
skewed distributions with a shift to higher values, as expected from the non-linear 
formation probability. The scaled higher moments: the variance 
$\sigma^2/\langle n_d \rangle$, the skewness $S\sigma$ and the kurtosis 
$\kappa\sigma^2$ all differ significantly from the Poisson expectation of unity.
The departure from Poissonian distribution is larger if proton and neutron number 
fluctuate together (Model A), but also independent proton and neutron 
fluctuations (Model B)  lead to clearly non-Poissonian shape.

\begin{figure}[t]
\includegraphics[width=0.5\textwidth]{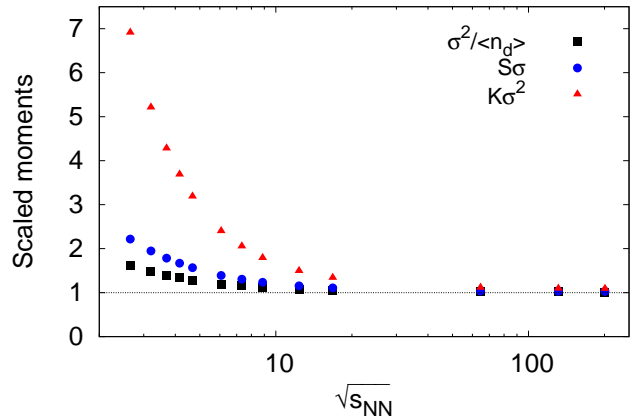}
\caption{\label{fig:mom}  The energy dependence of the moments $\sigma^2/\langle n_d \rangle$, $s\sigma$, and $k\sigma^2$ of the deuteron distribution 
obtained from Model A compared to the Poisson expectation}
\end{figure}

\begin{figure}[t]
\includegraphics[width=0.5\textwidth]{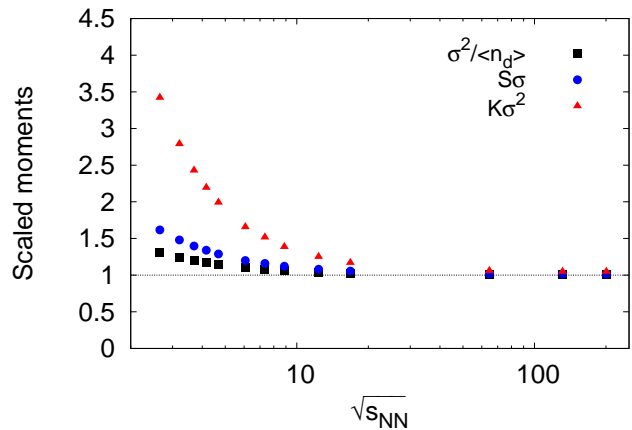}
\caption{\label{fig:mom2}  The energy dependence of the moments $\sigma^2/\langle n_d \rangle$, $s\sigma$, and $k\sigma^2$ of the deuteron distribution in the coalescence model assuming indepenedent proton and neutron fluctuations (Model B) compared to the Poisson expectation.}
\end{figure}

Next we explore the energy dependence of the moments of the deuteron distribution and compare to the Poisson expectations. Figures ~\ref{fig:mom} and \ref{fig:mom2} 
show 
the scaled moments $\sigma^2/\langle n_d \rangle$, $S\sigma$, and $\kappa\sigma^2$ as functions of collision energy for Models A and B, respectively. 
We observe a clear deviation from the Poisson expectation for all the higher moments. The deviation is very strong at low energies, where both coalescence parameter $B$ and the mean proton and neutron numbers are large, 
which results in sizeable fluctuations of the mean of Poissonian deuteron number distribution
given by eq.~(\ref{coal}) or eq.~(\ref{e:lamb}). This leads to  
even larger fluctuations of the deuteron number. The effect could be possibly observed for energies up to about 5 GeV in all the moments and even up to higher energies in kurtosis only.

Removing the correlation between initial proton and neutron 
fluctuations clearly weakens the effect, as can be seen in Fig.~\ref{fig:mom2}. 
The scaled moments attain approximately one half of the values obtained for 
neutron number  coupled to the proton number. We can still conclude, however, that 
we should be able to observe the deviation of the moments of the deuteron multiplicity distribution in case of coalescence regardless of the strength of the correlation of nucleon fluctuations. Additional information about the degree of this correlation can be provided by proton-deuteron multiplicity correlations, as shown in the Appendix.


\section{Summary}

In the light of the ongoing debate about 
the origin of the deuterons we propose to  measure data on fluctuations of deuterons. 
Employing standard approach and using the deuteron formation probability as 
proportional to the square of the nucleon yield we obtain strongly non-Poissionian 
distribution of the deuteron yield. 
Exact shape of deuteron distribution depends on whether proton and neutron 
yield are correlated or not. We tested both extremes---i.e., strongly correlated 
and completely independent---and the departure from Poissonian is always large. 
This allows to disentangle the direct grand canonical production of deuterons 
(and other clusters) from the formation of deuterons by coalescence. 

For simplicity, we have assumed that the initial proton number follows a
Poisson distribution. This is fine for measurements with small acceptance. 
If baryon number is exactly conserved within the acceptance 
window, then  the initial proton number should
follow binomial distribution. Nevertheless,  all our results should not change qualitatively.
Note also, that direct comparison with experimental data will also have to
include the fluctuations of volume. 

Our predictions 
are testable by the current experiments at the RHIC-BES and later by FAIR.

	
\begin{acknowledgments}
We thank Igor Mishustin for enlightening discussions. 
The computational resources were provided by the LOEWE Center for Scientific Computing (LOEWE-CSC) of the University of Frankfurt, Germany. 
This work was financially supported by the Helmholtz International Center for FAIR (HIC for FAIR) within the Hessian LOEWE initiative.
ZF and BT acknowledge partial support by grant No's 
APVV-0050-11, VEGA 1/0469/15 (Slovakia) and BT acknowledges 
M\v{S}MT grants LG13031 (Czech Republic).
ZF thanks the DAAD for the support during the stay at the 
Frankfurt Institute for Advanced Studies. 
\end{acknowledgments}


\appendix*
\section{Correlations}

Another observable which distinguishes thermal grand-canonical production and 
cluster production via coalescence is the correlation of proton and deuteron multiplicities. 
In the grand-canonical statistical approach the proton and deuteron multiplicities both 
fluctuate independently according to the Poisson distribution. 
However, in the coalescence scenario the deuteron fluctuations are connected to the initial 
proton and/or neutron number and thus to the observed nucleon number fluctuations, as well.
This leads to positive correlation between proton and deuteron multiplicities. 
On the other hand at a fixed initial proton number $n_i$ a larger deuteron multiplicity $n_d$ results in a smaller final proton number $n_p = n_i - n_d$, 
which introduces anticorrelation between $n_p$ and $n_d$. 

To explore to what degree the proton and deuteron multiplicities are correlated we evaluate 
the correlation coefficient $\rho$ defined as
\begin{equation}
\label{corr}
\rho(n_p,n_d) = \frac{\sum\limits_{k}(n_{p_k}-\lambda_p)(n_{d_k}-\lambda_d)}{\sigma_p\sigma_d},
\end{equation}
where $\lambda_p$($\lambda_d$) and $\sigma_p$($\sigma_d$) are the mean value and width of proton (deuteron) multiplicity distribution. The mean value and the width for both protons and deuterons are calculated using the distributions we derived earlier. 

Now we have to distinguish between the two coalescence models we introduced. 
First, we  explore Model A with strongly correlated initial proton and neutron fluctuations. 
To calculate the correlation coefficient consistenly within this approach we sum over all 
possible $(n_i,n_d)$ states:
\begin{equation}
\label{corr:coal}
\rho(n_p,n_d) = \frac{\sum\limits_{n_i,n_d}P(n_i)P(n_d|n_i)(n_i-n_d-\lambda_p)(n_d-\lambda_d)}{\sigma_p\sigma_d}.
\end{equation}  
\begin{figure}[t]
\includegraphics[width=0.5\textwidth]{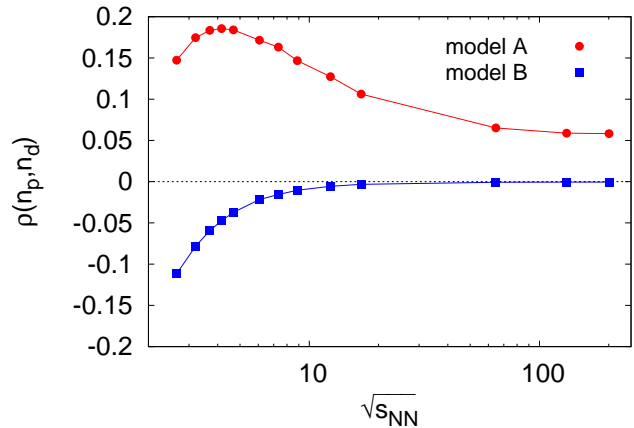}
\caption{\label{fig:corr}  Energy dependence of the correlation coefficients between proton number and deuteron number in the case of coalescence with strongly correlated proton and neutron fluctuations (Model A) and in the case of coalescence with independent proton and neutron fluctuations (Model B).}
\end{figure} 
In Fig. ~\ref{fig:corr} the correlation paramater $\rho(n_p,n_d)$ is shown as function of energy. The proton and deuteron multiplicity are positively correlated. The correlation is stronger for lower energies, where the coalescence parameter is larger. An interesting feature of the results is the decrease of the correlation for the lowest energies. This effect is caused by the wide deuteron distribution which emphasizes the extreme low and high values of deuteron number leading to stronger anticorrelation. Overall correlation is thus reduced.

Next we will investigate the correlations in the coalescence model with independent initial nucleon fluctuations (Model B). 
In this case, we calcute the correlation coefficient by summing over all possible $(n_i,n_d,n_j)$ states:
\begin{multline}
\label{corr:coal2}
\rho(n_p,n_d) = \frac{1}{\sigma_p\sigma_d}\\
\times
\sum\limits_{n_i,n_d,n_j}P(n_i)P(n_j)P(n_d|n_i,n_j)\\
\times 
(n_i-n_d-\lambda_p)(n_d-\lambda_d).
\end{multline}   
The correlation parameter $\rho(n_p,n_d)$ calculated assuming independent proton and 
neutron fluctuations is shown in Fig.~\ref{fig:corr} as function of energy and compared
with the result of Model A. There is a 
marked difference between the two models. 
The proton and deuteron multiplicities are now anticorrelated. 
The anticorrelation is caused by an interplay of two effects. Firstly, the number of deuterons now depends only linearly on the number of protons, as compared to the squared proton yield in 
Model A. Thus, the correlation is weaker. On the other hand, the anticorrelation is reinforced by the neutrons fluctuating independently, that cause an increase of higher deuteron production when the proton yield is lower, but neutron yield is high and vice-versa. 

The proton-deuteron multiplicity correlation measurement yields very different results for the two coalescence scenarios. However, in both scenarios we see a clear deviation from a Poisson-like uncorrelated thermal proton and deuteron production. We propose this measurement as complementary to the moments of the deuteron distribution that could possibly help to shed some light on the degree of correlation in the nucleon production.    


\end{document}